\documentstyle[epsf,twocolumn,fleqn]{article}

\begin{document}
\renewcommand{\textfraction}{0.1}
\renewcommand{\topfraction}{0.8}
\rule[-8mm]{0mm}{8mm}
\begin{minipage}[t]{16cm}
\begin{center}
{\Large \bf Spin correlation functions and N\'{e}el order in the 
2D Heisenberg model: Effects of spatial anisotropy\\[4mm]}
C.~Schindelin$^{\rm a}$, D. Ihle$^{\rm b}$, S.-L. Drechsler$^{\rm\, c}$ 
and H. Fehske$^{\rm a}$\\[3mm]
$^{\rm a}${Physikalisches Institut, Universit\"at Bayreuth, 
D-95440 Bayreuth, Germany}\\
$^{\rm b}${Institut f\"ur Theoretische Physik, Universit\"at Leipzig, 
D-04109 Leipzig, Germany}\\
$^{\rm c}${Institut f\"ur Festk\"orper- und Werkstofforschung Dresden e.V.
D-01171 Dresden, Germany}
\\[4.5mm]
\end{center}
{\bf Abstract}\\[0.2cm]
\hspace*{0.5cm}
The ground-state properties of the square-lattice spin-1/2 
Heisenberg antiferromagnet with spatially anisotropic 
couplings are investigated by Green's-function projection approaches.
The staggered magnetization and the two-spin correlators are calculated;
 the competition between magnetic long- and short-range order is discussed
in comparison with experiments on $\rm Sr_2[Ca_2]CuO_3$.\\[0.2cm]
\end{minipage}\\[4.5mm]
\normalsize
Motivated by experiments on quasi-1D quantum spin systems, such as 
$\rm Sr_2CuO_3$ and $\rm Ca_2CuO_3$~\cite{REHDM97}, many efforts were 
made to clarify the dimensional crossover in the square-lattice 
spin-1/2 antiferromagnetic (AFM) Heisenberg model~\cite{AGS94,Sa99,ISWF99} 
\begin{equation}
  {\cal H} = \frac{J_x}{2}
\Big[ \sum_{\langle i,j\rangle_x}{\bf S}_i{\bf S}_j 
+R  \sum_{\langle i,j\rangle_y}{\bf S}_i{\bf S}_j\Big] \,.
\label{model}
\end{equation}
Here $R=J_y/J_x$ (throughout we set $J_x=1$), and $\langle i,j\rangle_{x,y}$ 
denote nearest neighbors along the $x$-, $y$-directions. 
In the ground state, the staggered magnetization reveals
a transition from a long-range ordered (LRO) N\'eel state 
to a spin liquid with AFM short-range order (SRO) at the critical 
ratio $R_c$. Quantum Monte Carlo data provide strong evidence
for $R_c=0$~\cite{Sa99}, which also results from RPA
theories~\cite{MSS92,REHDM97} and (multi-) chain mean-field 
approaches~\cite{Sa99}. In previous work~\cite{ISWF99}, based on a
spin-rotation-invariant (SRI) Green's function theory and Lanczos 
diagonalizations, we found a sharp crossover in the spatial dependence
of the spin correlation functions at $R_0\simeq 0.2$.

In this paper we mainly focus on the SRO properties of the model~(\ref{model})
at $T=0$ studied by a generalized RPA theory compared with the SRI 
theory of Ref.~\cite{ISWF99}. Both approaches are based on the projection
method for two-time retarded Green's functions in calculating 
the dynamic spin susceptibility $\chi^{+-}({\bf q},\omega)=
-\langle\langle S_{\bf q}^+;S_{-{\bf q}}^-\rangle\rangle_{\omega}$. 
\begin{figure}[t]
\unitlength1mm
\begin{picture}(70,74)
\end{picture}
\end{figure}
First, we extend the non-SRI theory of Ref.~\cite{KS94} to the case $R\neq 1$,
 hereafter referred to as theory~I. Introducing two sublattices $(a,b)$
and taking the basis  ${\bf A}=(S_{\bf q}^{a+},S_{\bf q}^{b+})^T$,
where $S^+_{\bf q}=\frac{1}{\sqrt{2}}(S_{\bf q}^{a+}+S_{\bf q}^{b+})$, we get
\begin{equation}
\chi^{+-}({\bf q},\omega)=-\frac{M^{(1)}_{\bf q}}{\omega^2
-\omega_{\bf q}^2}\,,
\label{chi}
\end{equation}
with $M_{{\bf q}}^{(1)}=-4 C_{1,0}[1-\cos q_x+ R\zeta(1-\cos q_y)]$,
$\zeta=C_{0,1}/C_{1,0}$, $C_{\bf r}=
(C_{\bf r}^{+-}+2C_{\bf r}^{zz})/2\equiv C_{n,m}$,
$C_{\bf r}^{+-}
=\langle S_{\bf 0}^+S_{\bf r}^{-} \rangle
=\frac{1}{N}\sum_{\bf q} 
\frac{M_{\bf q}^{(1)}}{2 \omega_{\bf q}} \mbox{e}^{i{\bf q}{\bf r}}$,
and ${\bf r}=n {\bf e_x} + m {\bf e_y}$. 
The magnon spectrum $\bar{\omega}_{\bf q}=\omega_{\bf q}/Z_c^x$ is
\begin{equation}
\bar{\omega}_{\bf q}^2=
(1+R \zeta)^2 -(\cos q_x+R \zeta \cos q_y)^2\,,
\label{md}
\end{equation}
where $Z_c^x=-2 C_{1,0}/\langle S^{az}\rangle$. Using the identity  
$S_i^z=-\frac{1}{2}+S_i^+S_i^-$, the sublattice magnetization 
$m\equiv \langle S^{az}\rangle = - \langle S^{bz}\rangle $ is given by 
\begin{equation}
m=\Big[\frac{2 (1+R\zeta )}{N} \sum_{\bf q} \bar{\omega}_{\bf q}^{-1} 
\Big]^{-1}\,.
\label{mag}
\end{equation}
The theory contains one free  parameter $\zeta$ which we fix by the 
requirement $C_{0,1}^{+-}/C_{1,0}^{+-}\stackrel{(!)}{=}\zeta$. 
The spin-wave velocity renormalization factor $Z_c^x$ is calculated from 
$Z_c^x=(1+R)\Big[\frac{1}{N}\sum_{\bf q} \bar{\omega}_{\bf q} \Big]^{-1}$. 
In the RPA theory of Ref.~\cite{MSS92}, $m$ is given by Eqs.~(\ref{mag}) 
and~(\ref{md}) with $\zeta\equiv 1$. For $R\ll 1$ we have
$m=0.303[1-0.386\ln(R\zeta )]^{-1}$~\cite{REHDM97}.

In the SRI theory~\cite{ISWF99}, hereafter referred to as theory~II, the 
basis is chosen as ${{\bf A}=(S_{\bf q}^+, i \dot{S}_{\bf q}^+)^T}$ 
yielding $\chi^{+-}({\bf q}, \omega)$ and $M_{\bf q}^{(1)}$ given by 
Eq.~(\ref{chi}). The spectrum is calculated in the approximation 
$-\ddot{S}^+_{\bf q}=\omega_{\bf q}^2 S^+_{\bf q}$, where 
$\omega_{\bf q}$ is expressed by correlation functions 
and vertex parameters. Again, one parameter is free and may
be determined by adjusting either the ground-state energy per site
$\varepsilon(R)$~\cite{AGS94} (case A), the uniform susceptibility (case B), 
or $m(R)$~\cite{Sa99} (case C) [ for details see Ref.~\cite{ISWF99}].

As seen in Fig.~1, the LRO in theory~I is reduced compared with the 
RPA result~\cite{MSS92} due to the ratio~$\zeta$ expressing the SRO
anisotropy. Considering, e.g., the ordered moment in $\rm Ca_2CuO_3$,
where $R=0.023$~\cite{REHDM97}, in theory~I we get $\zeta=0.157$ and
$m=0.0956$ exceeding the experimental value~\cite{REHDM97} by a factor of 
about two, whereas the RPA and chain mean-field theory
($m=0.72 \sqrt{R}$ at $R\ll 1$~\cite{Sa99}) yield $m=0.123$ and
$m=0.109$, respectively. Comparing $\varepsilon (R)$ (inset) with the 
Ising expansion data of Ref.~\cite{AGS94}, theories~I and~II~C 
(input $m(R)$ of Ref.~\cite{Sa99}, cf. Fig.~1) yield insufficient
results at $R\stackrel{<}{\sim} 0.25$. On the other hand, in theory~II~B
($R_c\simeq 0.24$~\cite{ISWF99}), $\varepsilon(R)$ nearly agrees with 
the exact data at $R\stackrel{<}{\sim} 0.25$. 
That is, in the Green's-function theories describing LRO with
$R_c=0$, the SRO at $R\stackrel{<}{\sim} 0.25$ is reproduced inadequately. 
\begin{figure}[h]
\centerline{\mbox{\epsfxsize 8.5cm\epsffile{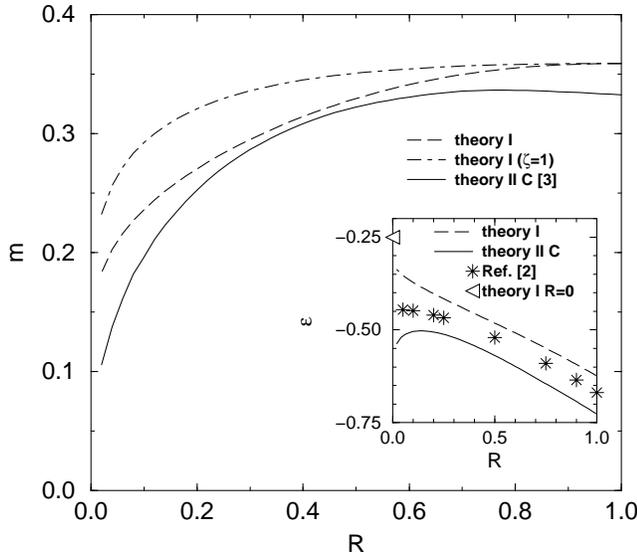}}}\vspace*{-0.5cm}
\caption{\small R-dependence of the magnetization $m$
and of the ground-state energy per site $\varepsilon$ (inset).}
\end{figure} 
The same qualitative behavior can be seen from $C_{\bf r}$ depicted in
Fig.~2. Compared with theory~II~A ($R_c\simeq 0.24$), where the correlators
reasonably agree with the exact diagonalization data~\cite{ISWF99},
theory~I becomes unsatisfactory at $R\stackrel{<}{\sim} 0.25$. There we have
$2|C_{1,0}^{zz}|\ll |C_{1,0}^{+-}|$ and, for $R<0.1$,
$C_{1,0}^{zz}>0$ being incompatible with the AFM SRO. 
Equally, the correlators $C_{1,1}$ and $C_{2,0}$ (inset) 
in theory~II~C strongly deviate from those in theory~II~A
at $R\stackrel{<}{\sim} 0.25$.

To conclude, our results call for an improved theory which
may describe both the LRO and SRO equally well and explain
the very small moments in $\rm Sr_2[Ca_2]CuO_3$. 
\begin{figure}[h]
\centerline{\mbox{\epsfxsize 8.5cm\epsffile{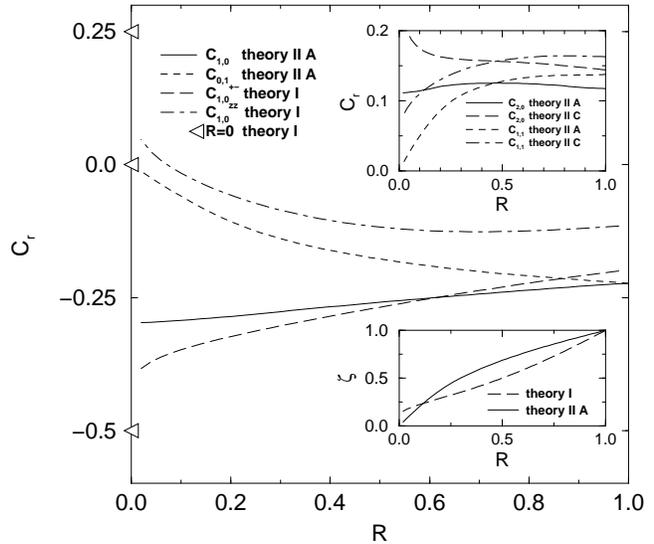}}}\vspace*{-0.5cm}
\caption{\small Nearest-neighbor and longer ranged (upper inset) 
spin correlation functions versus $R$. The lower inset demonstrates
that there is no decoupling transition; i.e. $\zeta>0$ $\forall R$,
\mbox{contrary to the suggestion in~\cite{PSZ93}}.}
\end{figure}\end{document}